\documentclass[iop]{emulateapj}
\usepackage{natbib}
\usepackage{hyperref}
\usepackage{amsmath}
\usepackage{graphicx}
\usepackage{verbatim}
\usepackage{epsfig}
\usepackage{xcolor}
\usepackage{ulem}

\shorttitle{Formation of hydroxylamine }

\begin{document}

\title{Formation of hydroxylamine on dust grains via ammonia oxidation}
\author{Jiao He, Gianfranco Vidali}

\affil{Physics Department, Syracuse University, Syracuse, NY 13244, USA}
    \email{gvidali@syr.edu}
\author{Jean-Louis Lemaire}
\affil{Paris Observatory, Paris, France}
\author{Robin T. Garrod}
\affil{Center for Radiophysics and Space Research, Cornell University, Ithaca, NY 14853, USA}
\begin{abstract}
The quest to detect prebiotic molecules in space, notably amino acids, requires an understanding of the chemistry involving nitrogen atoms. Hydroxylamine (NH$_2$OH) is considered a precursor to the
amino acid glycine. Although not yet detected, NH$_2$OH is considered a likely target of detection with ALMA. We report on an experimental investigation of the formation of hydroxylamine on an amorphous silicate surface via the oxidation of ammonia. The experimental data are then fed into a simulation of the formation of NH$_2$OH in dense cloud conditions. On ices at 14 K and with a modest activation energy barrier, NH$_2$OH is found to be formed with an abundance that never falls below a factor 10 with respect to NH$_3$. Suggestions of conditions for future observations are provided.

\end{abstract}

\section{Introduction}
With  molecules detected in space  increasing in number and growing in complexity, one of the main goals in astronomy today is to find a link between astrochemistry and astrobiology \citep{Wincel2000}. Prebiotic molecules have been already observed that can be considered as precursors of complex organic compounds like amino acids and prebiotic material. N-bearing molecules are necessary precursors of such prebiotic material and among them hydroxylamine (NH$_2$OH) appears to be one of the best candidates. Both experimental and theoretical studies \citep{Blagojevic2003,Snow2007} have shown that in its protonated form it can react in the gas phase with acetic and propanoic acids to yield protonated glycine (the simplest non-chiral amino-acid) and  protonated alanine (the second simplest and chiral
amino-acid). However, the question is still under debate whether such a synthesis happens in the gas phase or by surface reactions. Actually, in a recent work \citet{Barrientos2012} dismiss the possibility that
precursors of glycine could be efficiently produced from
the reactions of hydroxylamine-derived ions with acetic
acid  in the gas phase. This is confirmed by recent work \citep{Garrod2013} using a new three-phase chemical model that fully couples gas-phase, grain-surface and bulk-ice chemistry, which shows that glycine can be formed much more efficiently on grain surfaces than through gas-phase reactions.

Currently, the main problem  is  the lack of astronomical detection (either in the inter- or circum-stellar media, ISM and CSM) of both hydroxylamine  and  glycine (NH$_2$CH$_2$COOH), and this is  obviously correlated to their low predicted abundances. Searches for hydroxylamine have been done by \cite{Pulliam2012} observing seven bright sources toward IRC+10216, Orion KL, Orion S, Sgr B2(N), Sgr B2(OH), W3IRS5, and W51M using the NRAO 12m telescope on Kitt Peak, and by \citet{Charness2013} toward L1157-B1 using the CARMA Radio Telescope Array.  Glycine of cometary origin has been detected in STARDUST-returned samples \citep{Glavin2008,Elsila2009}, but it is unknown whether it represents pristine interstellar material. A chemically related species that is a possible direct precursor of glycine, amino acetonitrile (NH$_2$CH$_2$CN), has been detected \citep{Belloche2008} in Sgr B2(N). Glycine detection in the ISM or CSM is claimed to be highly plausible using ALMA towards bright nearby sources with narrow emission lines \citep{Garrod2013}. Such sources are luminous high-mass young stellar objects and low-mass protostars, also called hot cores and corinos. They are regions surrounded by dense material (10$^{8}$ cm$^{-3}$ for hot cores and 10$^{6}$ cm$^{-3}$ for hot corinos) and warm ($\sim$250 K and 100 K, respectively)  close to the core. This material is  much colder farther away. Deep within such dense regions, cosmic-ray (CR) ionized helium, He$^+$, breaks up  the very stable N$_2$ molecule, liberating  N atoms that interact with abundant H atoms to yield NH$_3$, which then sticks to cold grains \citep{Tielens2005}. It is worth noting that the composition of hot-cores resembles that of interstellar ices with high abundances of hydrogenated species as H$_2$O and NH$_3$ \citep{Tielens2005}. Hot cores and corinos are in general places where many terrestrial-like organic molecules are observed. More recently, radiative transfer calculations of  glycine emission show that low-mass star-forming regions, in their earliest stages of cold pre-stellar cores, may be better regions for the detection of glycine \citep{Jimenez-Serra2014}. Water detection in the low-mass pre-stellar core L1544 \citep{Caselli2012} indicates that a fraction of the grain mantles has desorbed into the gas phase. This makes the observation of glycine plausible, assuming it co-desorbs with water. After the first detection of nitrogen hydrides in Sgr B2 \citep{Goicoechea2004}, a complex region encompassing a variety of physical conditions and in particular shocks, the new detection with Herschel/HIFI of NH (and ND), NH$_2$, and NH$_3$ towards the external envelope of the protostar IRAS 16293-2422 \citep{Hily-Blant2010,Bacmann2010} has not improved our knowledge of the interstellar chemistry of nitrogen hydrides. In a recent publication, \citet{LeGal2014} develop a new model that could explain the formation of these hydrides in the gas phase while some of the modelers mentioned above claim surface formation. Concerning oxygen, one of the conclusions of deep observations of O$_2$ towards NGC 1333 IRAS 4A protostar \citep{Yildiz2013} is that the observed low molecular oxygen abundance is due to the freeze-out of atomic O onto grains. All these observations lead to the conclusion that on the grains in dense clouds surrounding protostellar cores, there can be NH$_3$ and O  present.

On the experimental side, two important facts are of interest. First, concerning NH$_3$, it has been experimentally demonstrated \citep{Hidaka2011} that the successive hydrogenation of N atoms trapped in a solid N$_2$ matrix at 10 K leads to NH$_3$ formation. Second, the formation of NH$_2$OH has been observed after irradiating an ammonia-water ice at 90-130 K with UV photons \citep{Nishi1984}, and at 10 and 50 K with 5 keV electrons \citep{Zheng2010}. The chemical processing of H$_2$O-NH$_3$ ices induced by heavy ions \citep{Bordalo2013} leads also, by radiolysis, to the formation of N$_2$O, NO, NO$_2$, and NH$_2$OH. As mentioned above, \citep{Blagojevic2003} showed that the reaction of hydroxylamine ions, NH$_2$OH$^+$, with acetic or propanoic acid makes ionized glycine and alanine. Much earlier, from about 1960 to mid 1980s, the NH$_3$+O reaction has been the subject of numerous experiments reviewed by \citep{Cohen1987}. They have been performed at moderate temperature (450 to 850 K) in flowing and static systems \citep{Perry1984,Baulch1984} where ground state O($^3$P) is obtained by laser photolysis, or at higher temperatures (850 to 2200 K) in flames \citep{Fenimore1961} or in shock tube experiments \citep{Fujii1986}. Even if such a reaction has never been studied at very low temperature, it has been proposed that excited NH$_3$O intermediate could rearrange to give stable hydroxylamine NH$_2$OH \citep{Baulch1984}. The formation of complex organic molecules involving atomic addition reactions on the cold surface of ices-covered interstellar grains in diverse environments, has been extensively reviewed recently \citep{Herbst2009}. As NO has been detected in the gas phase towards a
few high-mass star-forming regions, it has been suggested that hydroxylamine could be formed in the gas phase through non-energetic hydrogenation of NO ice under dark cloud conditions \citep{Charnley2001}. But recently, an NO ice hydrogenation surface reaction at low temperature has shown that an efficient route to NH$_2$OH formation was also possible with interstellar relevant ices \citep{Congiu2012a,Congiu2012b,Fedoseev2012}. It takes place at 10 to 15 K at both NO submonolayer and multilayer coverage. Such a process occurs without external energetic input but implies several successive steps of hydrogenation. It is worth mentioning a crossed molecular beam study of the O($^1$D)+NH$_3$ reaction at 295 K \citep{Shu2001}   forming vibrationally excited NH$_2$OH with a subsequent dissociation into two different reaction channels: OH+NH$_2$ and NHOH/NH$_2$O+H with 90\% and 10\% yield respectively. This reaction has been also studied theoretically \citep{Wang2004}. Finally, very recently,
hydroxylamine have been formed in a two-step
mechanism through the reaction of ammonia with hydroxyl
radicals in a Neon matrix at 4 K \citep{Zins2014}.

However, most of the objects mentioned above, if not all, either in high-mass or low-mass star-forming regions as well as in proto-stellar cores, share the common characteristic of having a large abundance of NH$_3$ (and also H$_2$O), in any case far higher than the NO abundance. This latter molecule appears to be present in a restricted number of objects relating to high-mass star-forming regions. The abundance of  NH$_3$  is asserted in observations of low-mass young stellar objects \citep{Bottinelli2010}  and even dark clouds (as the Bok globule Barnard 68 \citep{DiFrancesco2002}) and comets (as very recently detected on 67P/C-G by the ROSINA instrument aboard ROSETTA \citep{Altwegg2014}). Then the formation process in star-forming regions of hydroxylamine we present in this paper is a very likely more direct (single step) and realistic (considering the species involved) mechanism, compared to the one already known \citep{Congiu2012b}.

Actually this process is  well suited with respect to to hot-cores and protostellar cores. In addition to N resulting from N$_2$ dissociation under cosmic-rays irradiation, atomic O is also present for the same reason. Both N and O stick on the colder bare or ices covered grains, the ones deep into the cloud surrounding the core (A$_v$ $\sim$ 15 - 20). A very simplistic scheme could then be the following: (i) Very cold grains are covered by atoms (H, O, and N), either on the surface or in bulk, (ii) when the temperature begins to increase ($>$5-10 K), H atoms become mobile and  form NH$_3$ (and H$_2$O) by hydrogenation surface reactions , (iii) as the temperature continues to increase ($>$14 K in interstellar space conditions - corresponding in laboratory experiments to $>$20-25 K)), O atoms become mobile to finally react with NH$_3$ (and H$_2$O) already formed to to produce in particular NH$_2$OH through an insertion mechanism.

In this paper we present experimental evidence of the formation process of hydroxylamine via oxidation of NH$_3$ on dust grain analogs (Section 3). The experimental findings are then used in a 3-phase gas-grain model of a dark cloud (Section 4). A comparison of NH$_2$OH formation mechanisms and a discussion about suitable environments where NH$_2$OH could be observed are presented in Sections 5 and 6. 

\section{Experimental Setup}

\subsection{Apparatus}
The apparatus was described elsewhere \citep{He2011,Jing2013}; here we summarize the main features that are important for this study. The experiments were carried out in an ultra-high vacuum main chamber connected to an atomic/molecular beam line. The main chamber is pumped by a combination of a cryopump, turbomolecular pump, and ion pump and can reach $1.2\times10^{-10}$ torr after a bake-out. At the center of the chamber there is a 1 $\mu$m thick amorphous silicate thin film sample grown on a 1 cm$^2$ gold plated copper disk by electron beam physical vapor deposition. The detailed preparation and characterization of the sample can be found in \citet{Jing2013}.  The sample can be cooled to 8 K by a liquid helium-cooled sample holder and be heated up to 450 K by a cartridge heater behind the sample. At the beginning of each day, the sample is heated up to 400 K to clean it before cooling it down. The chamber pressure after cooling down is about $5\times10^{-11}$ torr. During exposure of the sample to the atomic/molecular beam, the pressure increases to about $1\times10^{-10}$ torr. At this pressure, the amount of water that stick on the sample from the chamber background gas  is negligible. A triple-pass quadrupole mass spectrometer (QMS) is mounted on a rotary platform to record desorbed species from the sample surface or to measure the beam composition. The QMS is placed in a differentially pumped enclosure and is fitted with  a cone aimed at a sample. The distance between the tip of the cone and the sample can be changed by repositioning the sample via a XYZ sample holder manipulator. This arrangement prevents products loosely adsorbed on sample holder parts from reaching the QMS, thus reducing the background level and improving the signal/noise. A radio-frequency dissociation source is mounted in the first stage of the triple differentially pumped beam line. A mass flow controller (Alicat MCS-5SCCM) is used to ensure stable gas flow. 

\subsection{Beam flux calibration}
\begin{figure}
 \epsscale{1}
 \plotone{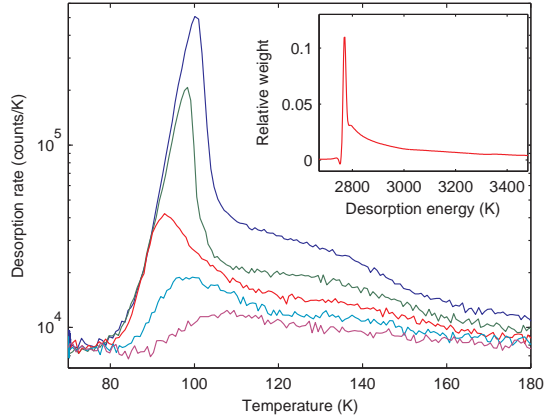}
\caption{TPDs with different exposure of NH$_3$ at 70 K. The exposure times are, from bottom to top, 0.5, 1, 2, 4, and 8 minutes,  respectively. Desorption energy distribution of NH$_3$ calculated from the 2 minutes exposure curve (1 ML deposition) is in the inset.}
\label{fig:NH3_TPD}
\end{figure}

The beam fluxes are calibrated by TPD experiments. With the the amorphous silicate sample kept at 70 K, ammonia gas is introduced from the molecular beam line. The gas flow is set to 0.3 sccm on the mass flow controller. After ammonia exposure, the sample is cooled down to below 40 K and then heated up at 1 K/s to about 200 K to do a TPD. In both  ammonia exposure stage and  TPD stage, the gas phase molecules are measured using the QMS. Since ammonia fragments in the QMS ionizer, both mass 17 amu (NH$_3^+$) and 16 amu (NH$_2^+$) are measured. The fragmentation of ammonia into NH$^+$/N$^+$ (mass 15/14 amu) is found to be negligible by direct beam measurements. A series of TPD experimental runs with different ammonia doses are performed. The TPD spectra are shown in Figure \ref{fig:NH3_TPD}. We calibrate the NH$_3$ coverage by analyzing the shape of the TPD profile. In the submonolayer region the TPD peak temperature should shift to lower values, because in the low coverage region molecules occupy the deep sites (higher desorption energies). Here a layer is actually an equivalent layer, since it is unclear whether ammonia would form clusters or islands on the surface; little experimental evidence support the existence of clusters or islands.  As the deep sites get filled up with increasing exposure, the TPD profile and peak temperature shift to lower temperatures \citep{He2014b}. In the multilayer region, the TPD spectra should show a common leading edge \citep{Kolasinski2008}, which is a typical first order desorption behavior. The trend of TPD traces shown in Figure \ref{fig:NH3_TPD} is in agreement with the one obtained by \citet{Bolina2005} for NH$_3$ desorption from a graphite surface. With the current experimental settings, 1 ML NH$_3$ coverage is achieved by 2 minutes exposure with 0.3 sccm gas flow. Thus the NH$_3$ flow is $0.5$ ML/minute, or equivalently, $5\times 10^{14}$ cm$^{-2}$minute$^{-1}$, assuming 1 ML $\sim$ $10^{15}\ \mathrm{cm}^{-2}$. Following the procedures as described in \citet{He2014b}, the desorption energy distribution of NH$_3$ from amorphous silicate surface can be obtained by direct inversion of the 1 ML trace in Figure \ref{fig:NH3_TPD}. The resulted distribution is shown in the inset of Figure \ref{fig:NH3_TPD}. Additional TPD runs are performed at a surface temperature of 10 K, 30 K, and 50 K. They show almost identical TPD curves as the one at 70 K. This suggests that the sticking of ammonia on the silicate surface at 70 K is unity, and the desorption rate at 70 K is negligible. 

In the atomic oxygen exposure, the O$_2$ gas flow passing through the flow controller is set to 0.1 sccm. At this flow the dissociation rate of O$_2$ is measured to be 42\%. The calibration of oxygen flux needs to be done differently from ammonia because O$_2$ is volatile and the sticking is not necessarily unity. The direct beam intensities of O$_2$ gas at both 0.3 sccm and 0.1 sccm are measured. Since it is known that 0.3 sccm corresponds to 0.5 ML/minute, the flow at 0.1 sccm can be calculated correspondingly. The O$_2$ flow at 0.1 sccm is about 0.34 ML/minute ($3.4\times 10^{14}\  \mathrm{cm}^{-2}\mathrm{minute}^{-1}$). When the RF is on, the beam intensities of O and O$_2$ are 0.29 ML/minute ($2.9\times10^{14}\ \mathrm{cm}^{-2}\mathrm{minute}^{-1}$) and 0.20 ML/minute ($2.0\times10^{14}\  \mathrm{cm}^{-2}\mathrm{minute}^{-1}$), respectively. With the radio frequency (RF) power on and the oxygen sent into the dissociation source, beam contamination is checked. The main contaminant is NO (mass 30 amu) due the small leak of air into the dissociation source. The NO (mass 30) signal is less than 3\% of the O signal, which is trivial in the context of the experiments performed here.

\label{sec:NH3_beam_calibration}

\subsection{Experimental procedures}
\begin{figure}
\epsscale{1}
\plotone{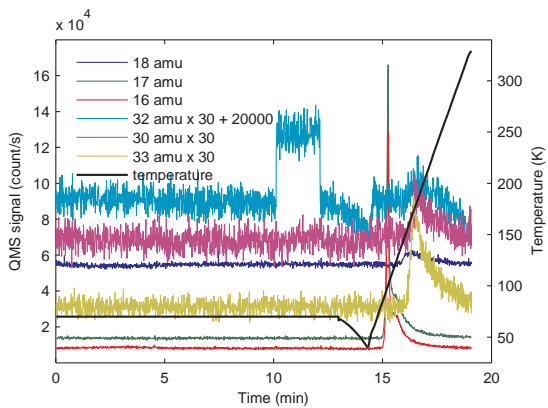}
\caption{A typical TPD of a NH$_3$ + O sequential exposure experiment at 70 K in which the QMS records simultaneously multiple signals. The temperature ramp (black line) shows that after the exposure is terminated, the sample is cooled to 40 K before the heating is started.}
\label{fig:NH2OH_example}
\end{figure}

The ammonia oxidation reactions were studied in ammonia and atomic oxygen sequential exposure TPD experiments. The surface was covered with ammonia before introducing atomic oxygen. Two ammonia coverages were used, 2 ML and 1/4 ML, representative of multilayer coverages and submonolayer coverages. After preparation of the ammonia sample, the residual ammonia in the beam line and dissociation source was cleaned  by flushing the beam line several times using oxygen. This ensures that almost no ammonia was mixed with oxygen.  The sample temperature in the exposure stage was chosen to be 70 K so that O/O$_2$/O$_3$ does not stick onto the surface while the sticking of NH$_3$ is unity. After the exposure stage, the surface was cooled down to below 40 K and then heated up linearly at 1 K/s to above 320 K to desorb species from the surface. The QMS recorded  simultaneously the signals of various masses during both exposure and TPD stages. Figure \ref{fig:NH2OH_example} shows the TPD spectra of a typical experimental run.

\section{Results and Analysis}

\subsection{Multilayer \texorpdfstring{NH$_3$}{NH3} + O}
In the NH$_3$ + O sequential exposure TPD experiments, mass 16, 17, 18, 30, 32, and 33 amu were measured by the QMS, see Figure \ref{fig:NH2OH_example}. Mass 16 amu is due to the fragmentation of NH$_3$ in the QMS ionizer; mass 17 amu is due to NH$_3$ and fragmentation of H$_2$O; mass 32 amu is due to O$_2$ and  fragments of O$_3$.  Mass 30 amu could be due to NO from the beam line (but the amount is small), or fragments of NH$_2$OH; mass 33 amu could be due to NH$_2$OH. HO$_2$ has the same mass but it is unlikely to be there because of the lack of detection of the products it fragments into. The mass 33 amu peak centered at around  180 K is accompanied by mass 30 amu and very small signals with masses 16, 17 and  32 amu at the same temperature; this suggests that the mass 33 amu peak at 180 K is due to NH$_2$OH. This agrees with the TPD peak attribution in \citet{Congiu2012a} (in the range of 160-200 K, peaking at T$\sim$ 190 K) obtained in an NO hydrogenation experiment and confirmed by infrared measurements (RAIRS). Mass 18 amu is due to water.

\begin{figure}
 \epsscale{1}
 \plotone{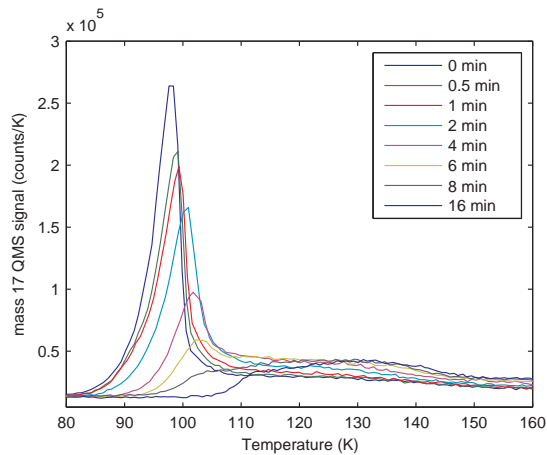}
\caption{Mass 17 amu QMS signal of the TPD from  an ice consisting of  NH$_3$ (4 min or 2 ML) followed by exposure to various doses of O at 70 K. }
\label{4min_NH3_m17}
\end{figure}

Figure  \ref{4min_NH3_m17} shows the mass 17 amu (NH$_3$) signal after depositing various doses of O on top of 2 ML (4 minutes exposure) NH$_3$. As the O exposure time increases, the NH$_3$ peak decreases and the peak position also shifts to higher temperatures. This is because the top layer of NH$_3$ is gradually converted to NH$_2$OH or other products, thus hindering the desorption of NH$_3$ molecules underneath. After about 6 minutes of O exposure, the first peak almost disappears, indicating the top layer of NH$_3$ is almost all converted to NH$_2$OH or other products. 

\begin{figure}
 \epsscale{1}
 \plotone{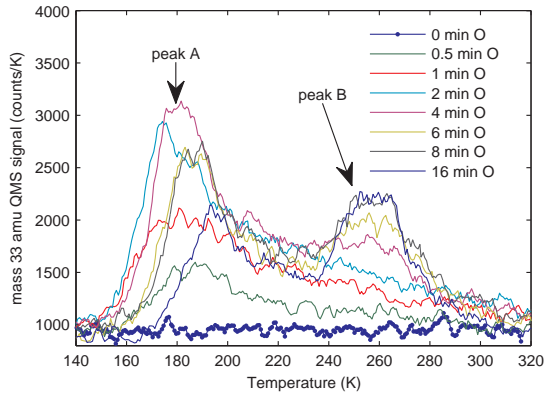}
\caption{As in Figure \ref{4min_NH3_m17} but for QMS mass 33 amu signal. }
\label{4min_NH3_m33}
\end{figure}

In Figure  \ref{4min_NH3_m33} two desorption peaks, peak A and peak B, are visible. Peak A shows up at O exposure as low as 0.5 minutes. This indicates that the ammonia oxidation reaction is efficient. As the O exposure increases from 0  to 4 minutes, the area of peak A increases at first, but then it decreases from 4 minutes to 16 minutes of O exposure. Peak A is followed, at larger O exposures, by peak B at $\sim$ 260 K. Notice that peak B starts to appear at 4 minutes of O exposure time at the expense of peak A. As mentioned above, peak A falls in the temperature range of the TPD mass 33 amu peak obtained by \citet{Congiu2012a}. Therefore we attribute peak A to the desorption of NH$_2$OH. Concerning peak B, however, in a subsequent
paper on an NO$_2$ hydrogenation experiment \citep{Ioppolo2014}, a mass 33 peak, confirmed by infrared
measurement spectra, appears at a temperature
T$\sim$ 245 K, close to the temperature of our second peak (peak B in our Figure \ref{4min_NH3_m33}).
Nevertheless, oddly there is no mention in their paper of
a peak A at 160 K.

\begin{figure*}
 \epsscale{1}
  \plottwo{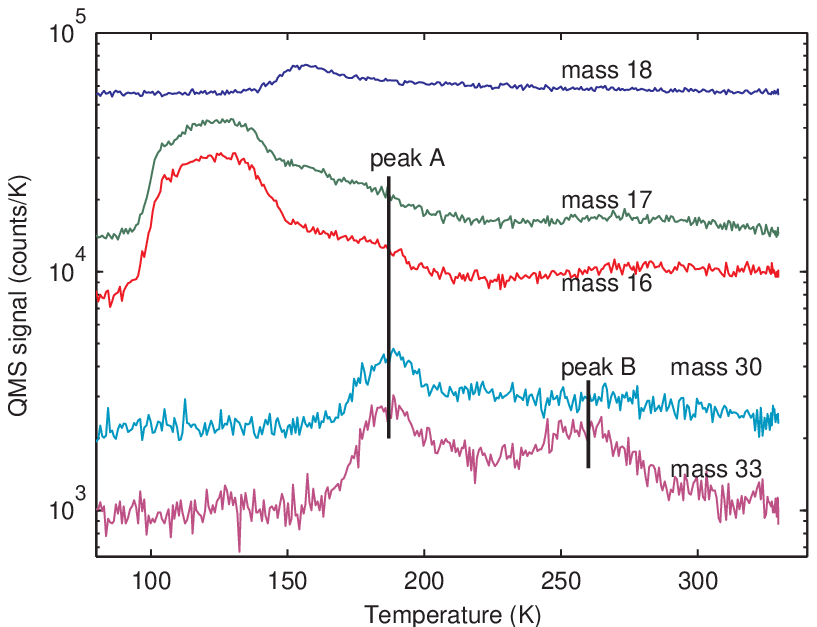}{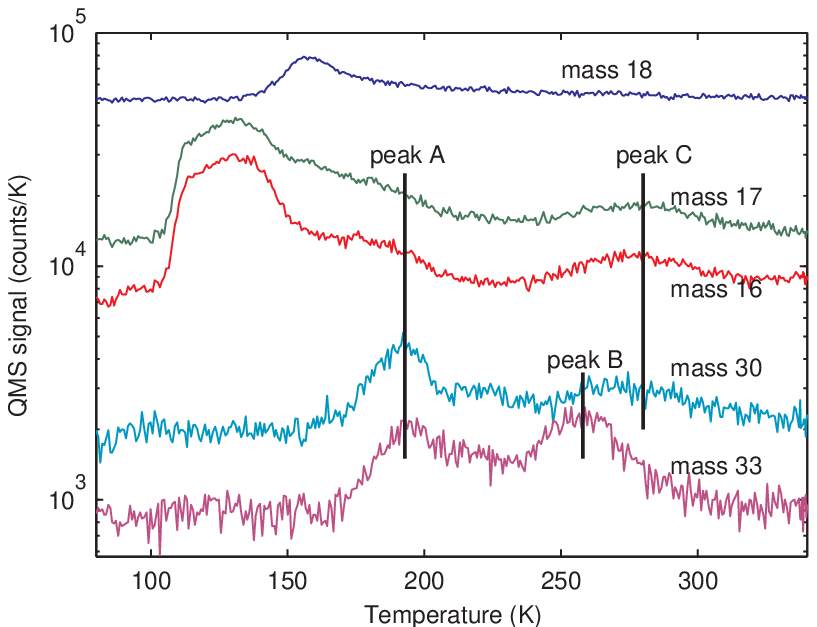}
   \caption{QMS signal of the TPD from an ice of NH$_3$ (4 min) + 8 (left) and 16 (right) minutes of O exposure. }
   \label{8and16}
  \end{figure*}

To find out what peak B represents, we show in Figure  \ref{8and16} a comparison of TPD traces for various masses for 2 ML of NH$_3$ + 8 or 16 minutes of O exposure, see left and right panels of Figure  \ref{8and16}, respectively. Peak A and peak B are marked by vertical lines. In both panels of Figure  \ref{8and16}, peak A (see trace of mass 33 amu) is accompanied by desorption of mass 30 amu, 17 amu, and 16 amu, while this is not true for peak B. This suggests that peak B is not due to NH$_2$OH. It could be the product of a fragmentation of a dimer \citep{delBene1972} or of an oxidation product of NH$_2$OH. At about 280 K, there is a peak C showing up for mass 30 amu, 17 amu, and 16 amu, but only when the O dose is high (16 minutes of O exposure), see right panel of Figure  \ref{8and16}. It could be due to a yet unidentified product formed in a further oxidation of NH$_2$OH. Peak B and peak C differs in both position and shape, therefore they should be attributed to different species.  

\subsection{Submonolayer \texorpdfstring{NH$_3$}{NH3} + O}
\begin{figure}
 \epsscale{1}
 \plotone{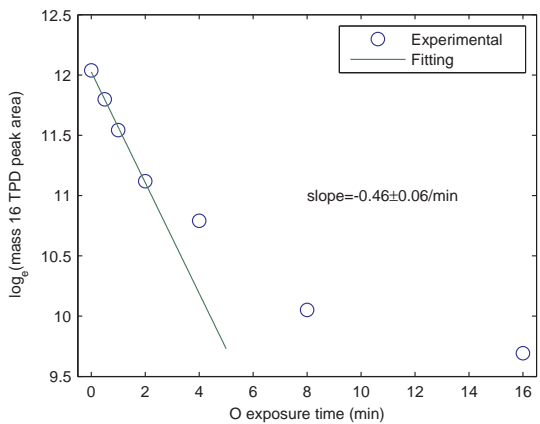}
\caption{Integrated TPD signs mass 16 amu from a deposit of  NH$_3$ (0.5 minutes or 1/4 of a ML) + various exposures of O at 70 K. }
\label{log}
\end{figure}

Experiments were also carried out at submonolayer NH$_3$ coverage. We exposed the silicate sample to NH$_3$ for  0.5 minutes, or  about 1/4 of a layer, with the same procedures as the ones followed to obtain the data in Figure \ref{fig:NH3_TPD}. This was followed by exposure to O.  Mass 16 amu was chosen to represent the NH$_3$ amount because the mass 17 amu signal has a non-trivial contribution from water fragmentation when the NH$_3$ signal is weak. In Figure \ref{log} we show the integrated mass 16 amu TPD trace for different exposures to O. With O exposure from 0  to 2 minutes, the amount of  NH$_3$ follows more or less an exponential decay. This is because when the O amount is small, oxidation dominates and the NH$_3$ destruction rate is proportional to the NH$_3$ amount. With a further increase in O exposure, the NH$_3$ decay does not follow a simple exponential decay anymore because of  possible secondary reactions. In Figure \ref{log}, a straight line is fitted to the log$_e$ plot from 0 to 2 minutes. The slope  is $-0.46\pm0.06\ \mathrm{minute}^{-1}$. In an exponential decay, the amount of NH$_3$ on the surface should follow $\exp(-\sigma\phi t)$, where $\sigma$ and $\phi$ are the reaction cross-section area and the beam flux, respectively. From it, we obtain the cross section $\sigma = -slope/\phi = (1.6\pm0.2)\times10^{-15}\ \mathrm{cm}^2$.

\subsection{Reaction energy barrier of \texorpdfstring{NH$_3$}{NH3} + O}
In typical ISM conditions, it is unlikely that situations arise similar to the ones that led to the appearance of the second TPD peak of mass 33 amu at high temperature attributed to a subsequent oxidation process. Therefore we focus on the first oxidation:
\begin{equation}
 \text{NH$_3$ + O $\rightarrow$ NH$_2$OH}
\end{equation}

When the surface is fully covered with NH$_3$ and the
surface temperature is 70 K, O could react with
NH$_3$ also via the Eley-Rideal or hot-atom mechanisms, i.e.  reaction without complete thermal accommodation of the oxygen atom.
Based on prior experiments \citep{He2014a}, we assume that at 70 K the residence time of O on NH$_3$ ice is negligible. We also assume that all the reactions take place during the exposure stage instead of the TPD stage, because at 70 K the mobility of oxygen should be relatively high. The degree of conversion of NH$_3$ can be gauged by looking at what remains of the NH$_3$ layers after various doses of oxygen. Figure \ref{4min_NH3_m17} shows the TPD mass 17 signal (NH$_3$) for various doses of O. The NH$_3$ dose is fixed at 4 minutes, which is equivalent to about $2\ ML$. As the O exposure time increases, the NH$_3$ peak decreases, and the peak position also shifts to higher temperatures. This is because the top layer of NH$_3$ is gradually converted to NH$_2$OH or other products, thus preventing NH$_3$ molecules underneath from desorbing. After about 6 minutes of O exposure, the first peak almost disappears, indicating that the top layer of NH$_3$ is almost all converted to NH$_2$OH or other products. Thus it takes at most 6 min of O ($1.74\ ML$) to convert $1\ ML$ of NH$_3$, suggestion an efficiency for the reaction of each oxygen atom of at least 0.575 (=1/1.74). 

Using this efficiency, an upper limit for the activation energy barrier, $E_\mathrm A$, for the O + NH$_3$ reaction may be estimated. 

We firstly assume that the incoming oxygen atom is thermalized on the surface, before overcoming the activation energy barrier to reaction. This thermalization is the most likely immediate outcome of the impact, except in a small minority of "direct-hit" trajectories, as the oxygen atom dissipates energy as it is drawn into the multi-dimensional potential of the surface. The atom is nevertheless available for immediate reaction if the activation energy barrier can then be overcome; in such a case, the reaction may be classed as Eley-Rideal, as it is not mediated by a diffusion process.

Assuming that the main loss mechanism for an oxygen atom on a pure NH$_3$ surface is either reaction with NH$_3$ or thermal desorption, the reaction efficiency may be formulated thus:

\begin{equation}
1 / 1.74 = \exp[-E_\mathrm{A} / T] / (\exp[-E_\mathrm{A} / T] + \exp[-E_\mathrm{des}(\mathrm O) / T])
\end{equation}

\noindent where $E_\mathrm{des}(O)$ is the desorption energy for an oxygen atom and T is the surface temperature. Adopting $E_\mathrm{des}$(O) = 1500 K and T = 70 K, this expression produces a value $E_{A}$(O+NH$_3$) = 1479 K.

This value constitutes an upper boundary of the reaction energy barrier because we assume a sticking coefficient of unity, and thence that each oxygen atom that reaches the surface participates in a reaction. If this is not the case, then the reaction energy barrier for the first oxidation should be even smaller. 

The above treatment does not consider the possibility that surface oxygen atoms could meet each other and react, reducing the efficiency of the reaction. Under conditions in which such reactions could be important, then in order for the O + NH$_3$ reaction to be efficient and the O + O reaction to be minimized, the reaction barrier for O + NH$_3$ would need to be lower than the oxygen diffusion barrier. This quantity has not been measured in the laboratory, but there are rules-of-thumb based on experimental data from other systems. The ratio of the diffusion energy barrier to the desorption energy is typically found to be $\sim$0.3 for crystalline surfaces \citep{Bruch2007}. For rough surfaces, values between 0.5 and 0.8 have been used \citep{Katz1999, Garrod2006}, although \citet{Garrod2011} found that values lower than $\sim$0.4 were most consistent with observed interstellar extinction thresholds for H$_2$O, CO and CO$_2$ ices. \citet{Garrod2013} adopted a value of 0.35, which would yield an upper limit on the O + NH$_3$ activation energy barrier as low as 525 K.

\subsection{Control experiments}
\begin{figure}
 \plotone{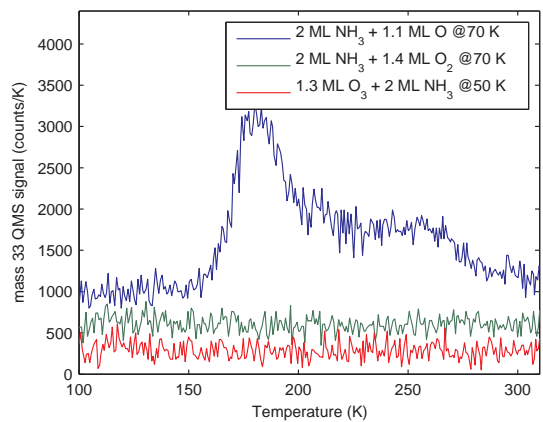}
\caption{Mass 33 QMS signal in the TPD from NH$_3$ + O$_2$ and O$_3$ + NH$_3$ deposited at 70K compared with that of NH$_3$ + O at 50 K. Curves are offset for clarity.}
\label{control}
\end{figure}

Control experiments were carried out to verify whether ammonia react with molecular oxygen or ozone. We exposed the ammonia layers to O$_2$ and found no mass 33 or 30 amu peaks in the TPD, signifying that there has been no conversion of ammonia to hydroxylamine in the presence of molecular oxygen. We also checked the reactivity of ozone with ammonia. Ozone was prepared on a clean silicate following the procedure described in \citet{He2014a}; then, 2 ML of ammonia were deposited on it. Again, there was no mass 33 or 30 amu peaks in the following TPD, see Figure \ref{control}. 
 
\section{Modeling}

In order to test the astrophysical importance of the O + NH$_3$ $\rightarrow$ NH$_2$OH reaction investigated here, we incorporate this surface mechanism into the recent three-phase gas-grain chemical kinetics model  MAGICKAL (Garrod 2013). The model parameters used here are intended to approximate those present under dark cloud conditions, during which significant ice mantles are formed on the surfaces of dust grains. The model utilizes the chemical network and initial elemental abundances used by \citet{Garrod2013}, assuming a generic, static dark-cloud gas density of $n_\mathrm H = 2 \times 10^{4}$ cm$^{-3}$ and visual extinction $A_\mathrm{V} = 10$. A gas temperature of 10 K is used in all model runs. We assume a binding energy for atomic oxygen of 1500 K,  based on an experiment that measured the desorption energy of O from compact amorphous water ice \citep{He2014c}. Diffusion barriers are set to a uniform fraction of binding energies for all species, $E_\mathrm{diff}$ = 0.35 $E_\mathrm{des}$, following \citep{Garrod2013}. Due to the low temperatures at which grain-surface ice mantles are formed, only surface chemistry is switched on in this model, although the three-phase treatment considers separate grain/ice surface and bulk mantle populations for all species (as well as gas-phase abundances). The three-phase model allows the composition of each layer within the ice mantle to be traced as it is deposited during the chemical evolution of the cloud (see \citet{Garrod2011}). Only Langmuir-Hinshelwood (i.e. diffusional) processes are considered in this model, as the Eley-Rideal mechanism tends to require surface coverages of close to unity (in models of astrophysical grains) to compete with L-H mediated reactions.

Reactions are also included in the network allowing NO molecules to be hydrogenated to NH$_2$OH by the sequential addition of atomic H. Following \citep{Garrod2013}, this includes a barrier-mediated reaction, H + HNO $\rightarrow$ HNOH, as well as an alternative H-abstraction branch, H + HNO $\rightarrow$ H$_2$ + NO, for which the barrier is assumed to be 1500 K in both cases. Following \citep{Hasegawa1992}, tunneling through activation energy barriers is treated using a rectangular barrier treatment, with a default width of 1 \AA. The faster of tunneling and thermal penetration is used, according to the grain temperature.

The model described above is used to produce a grid of models for a range of dust-grain temperatures from 10 -- 20 K, and using activation energy barriers, $E_\mathrm A$, for the O + NH$_3$ $\rightarrow$ NH$_2$OH reaction ranging from 0 -- 2500 K. A smaller set of models is also run to compare the effects of NH$_2$OH produced by this mechanism, with cases in which the barrier to the H + HNO $\rightarrow$ HNOH reaction is altered.

\subsection{Model results}

\begin{figure}
\epsscale{1.2}
 \plotone{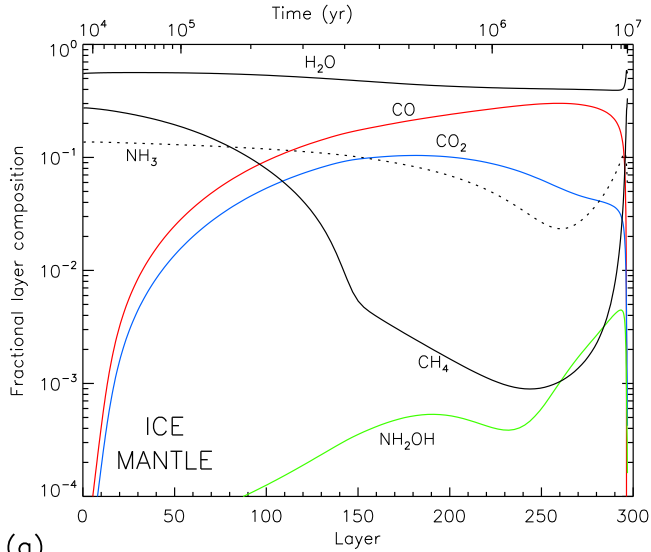}
 \plotone{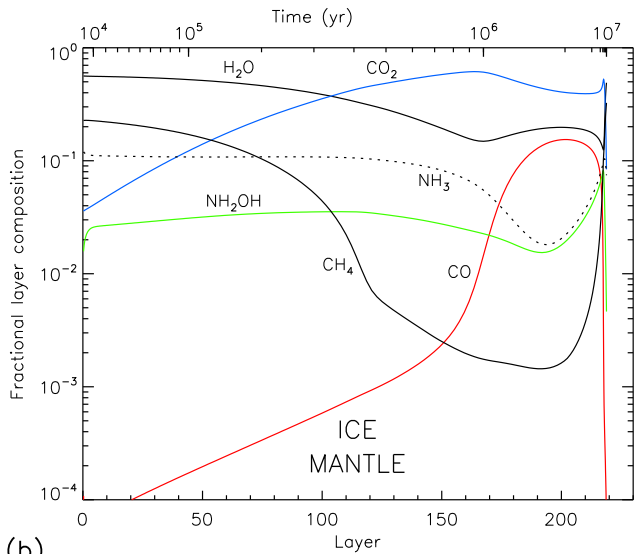}
 \plotone{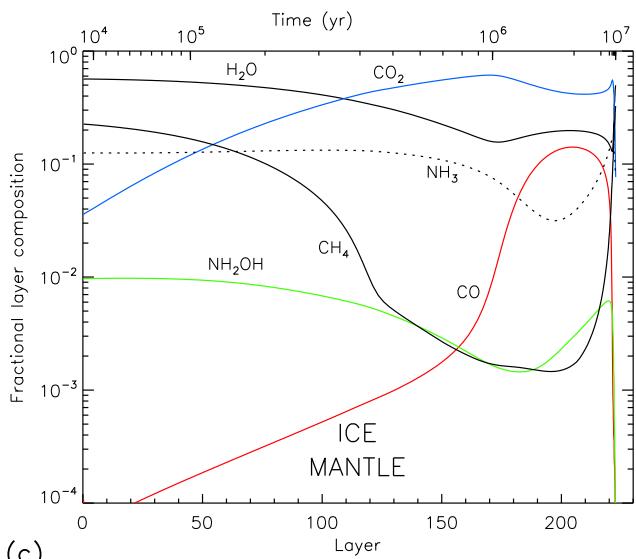}
\caption{Fractional composition of ices within each layer as a function of depth into the ice, lower horizontal scale, or, equivalently, as a function of age of the cloud, top horizontal scale. Panel (a) The activation energy barrier for the O+NH$_3$ reaction is 0~K and the temperature of the ice is 10~K. Panel (b) $E_\mathrm A$=1000~K and T=14~K. Panel (c) $E_\mathrm A$=2000~K and T=14~K.}
\label{RG1(a)}
\end{figure}

Figure \ref{RG1(a)}, panel (a), shows the composition of each layer in the grain mantle ice for the model run corresponding to $E_\mathrm A$=0 and T$_{\mathrm{grain}}$ = 10 K. Typically observed components are shown, along with NH$_2$OH. In this case, NH$_2$OH production is relatively small, and all such production is a result of the hydrogenation of NO. At 10 K, oxygen atoms are insufficiently mobile to be able to reach and react with NH$_3$ before they are hydrogenated to produce water, even under the assumption of a zero barrier for the O + NH$_3$ reaction. Here, the majority of NH$_2$OH  is produced via NO hydrogenation, which is strongly peaked at late time/outer layers.

Panels (b) and (c) show the results at 14 K, for activation energy barriers of $E_\mathrm A$=1000 and 2000 K, respectively. At this temperature, oxygen atoms are sufficiently mobile to meet ammonia molecules on the surface before being hydrogenated. In the $E_\mathrm A$=1000 K case, the reaction barrier is low enough (in comparison to the diffusion barrier for the oxygen atom to move away from the ammonia molecule) to allow reaction to occur with high efficiency. In this case, the majority of NH$_2$OH is formed through this reaction, and NH$_2$OH abundance closely follows that of NH$_3$, never falling below a factor of ten lower than NH$_3$ at any layer in the ice. NH$_2$OH maintains a fairly steady abundance throughout the ice, up to around 1 Myr (170 ML), but the NH$_2$OH production is enhanced at all times/depths. In the $E_\mathrm A$=2000 K case, the reaction barrier is too high to allow reaction before hydrogenation of oxygen occurs, making the NH$_3$ + O reaction inefficient; here NH$_2$OH production is again the result of NO hydrogenation.

\begin{figure}
\epsscale{1.2}
 \plotone{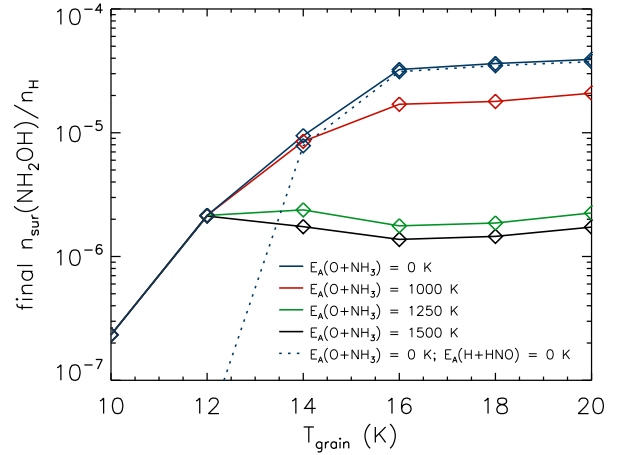}
\caption{Final model abundances of ice-mantle NH$_2$OH with respect to total hydrogen. The grain temperature is varied for a selection of activation-energy values for the NH$_3$ + O reaction. The dashed line shows the results assuming no barrier for either the NH$_3$ + $\rightarrow$ NH$_2$OH or H + HNOH $\rightarrow$ HNOH reactions.}
\label{RG3}
\end{figure}

Figure \ref{RG3} shows the final (maximum) grain-mantle abundances of NH$_2$OH as a function of total hydrogen in the cloud, for model runs assuming a range of grain temperatures from 10 -- 20 K, and activation energies for the O + NH$_3$ reaction ranging from 0 -- 1500 K. Results for the E$_\mathrm A$=500 K case are not shown, as they are identical to the $E_\mathrm A$ = 0 case; both values are sufficiently below the atomic oxygen diffusion barrier assumed in the model (525 K), such that the reaction with NH$_3$ occurs at an efficiency close to unity, if the temperature is high enough to allow the oxygen atom to diffuse and reach an NH$_3$ before it is hydrogenated and becomes immobile. This is the case for models of temperature 14 K and above; all models at 12 K or less produce the same results, with NH$_2$OH produced by NO hydrogenation.

Assuming a barrier of 1000 K, the reaction is still relatively efficient, falling by a factor $\sim$2 at grain temperatures of $\sim$16 K and above, compared to the $E_\mathrm A$ =0/500 K results. However, with a 1500 K barrier, reaction is inefficient, and yet higher barriers produce the same results; in such cases, NH$_2$OH is produced overwhelmingly by NO hydrogenation. A run with the O + NH$_3$ barrier set to 1250 K allows this reaction to become slightly dominant over the NO hydrogenation mechanism, producing a small, though significant, increase in NH$_2$OH production.

Models were also run to compare the case where the hydrogenation of NO by H atoms is assumed to occur either (i) without barriers or (ii) not to occur at all. Case (i), removal of the barrier to H + HNO $\rightarrow$ HNOH (while the alternative hydrogen-abstraction branch retains its barrier of 1500 K), in fact results in no appreciable difference from the runs that assume a barrier of 1500 K for H + HNO (the zero-barrier results are thus not plotted explicitly in the figure). NH$_2$OH production through this mechanism is therefore already at its highest efficiency. 

The alternative regime, case (ii), in which the H + HNO $\rightarrow$ HNOH barrier is increased to arbitrary height, such that the reaction does not occur, results in negligible NH$_2$OH production at 12 K and below, because all NH$_2$OH formation is dependent on the O + NH$_3$ reaction, which is inefficient at these temperatures. Above 12 K, NH$_2$OH is formed efficiently via O + NH$_3$, and there is only a small difference between these results and those in which the H + HNO reaction is allowed to occur at maximum efficiency.

\section{Discussion}
Experiments of oxygen exposure on submonolayer and multilayer ammonia ices show an efficient formation of NH$_2$OH with  a reaction energy barrier that can be as low as $\sim 525K$ and no higher than 1479K. The experiments were done on samples at 70 K to avoid contamination of molecular oxygen and ozone. At this temperature NH$_3$ is stuck on the surface while O has a short residence time. 
Experiments have been performed at both bilayer and submonolayer coverage. Simulations were done to find what rates of hydroxylamine formation one would obtain in simulated dense cloud conditions. 
According to the chemical models, at dust temperatures around 14 K or higher, the O + NH$_3$ reaction is found to be an efficient mechanism for the production of NH$_2$OH. While temperatures this high are not the typically considered dark-cloud dust temperatures, it is certain that the dust in dark clouds must pass through such temperatures on their way to the 8 -- 10 K that is more typically assumed. Furthermore, interstellar dust will again pass through the $\sim$14 K threshold at which this reaction becomes efficient as part of the evolution of hot cores.

\section{Astrophysical Implications}

NH$_3$ and H$_2$O have been detected in the gas phase around transient protostellar cores or inferred for O from the strong O$_2$ depletion. These cores are also classified as hot cores or hot corinos in the case of massive and medium/low mass nascent stars, respectively. These cores whose temperature can be as high as 100 K  close to the protostellar source (and even 250 K for the densest ones) are much colder far  from the core due to the strong extinction provided by the dusty molecular clumps in which they are embedded.  A possible scenario for the coldest regions where NH$_3$ and O are stuck on the grains and approaching  the core is that in a first stage (at temperature in the 14-50 K range) O is released in the gas phase and is then able to collide with NH$_3$ covered grains, inducing chemical reactions that can synthesize new species. These would later desorb from the grains when the core temperature increases. Alternatively, NH$_2$OH may be produced at lower temperatures, during the formation of the ice mantles, to be released at later times. (Models of such scenarios are beyond the scope of this primarily experimental study, but will be considered in future work).

Consequently, the present laboratory astrophysics experiment on O interaction with NH$_3$  should be useful to interpret the observational data.  
Then the question is of the non-observation of hydroxylamine while, at the same time, it has been demonstrated that it can be made in the laboratory in conditions almost similar to those in  protostellar cores.The upper limit relative abundance of NH$_2$OH with respect to H$_2$, as deduced from observations \citep{Pulliam2012} is found to be in the $3 \times 10^{-9}$ to $8 \times 10^{-12}$ range, depending on the object.  Interestingly, combining this result with the
relative abundance of NH$_3$ with respect to H$_2$ in a high-mass
star-forming region, 4.6 $\times$ 10$^{-8}$ \citep{Battersby2014} (a value of  2 $\times$ 10$^{-8}$ is already reported in
the case of a cold dense cloud like TMC-1 \citep{Turner2000}), we deduce an average fractional abundance of NH$_2$OH with respect to NH$_3$ in the 0.7 $\times$ 10$^{-1}$ to
2 $\times$ 10$^{-4}$ range, close to the values deduced from
the proposed model.
Several reasons may explain the non-observation: the yield of such reactions could be quite low due to the narrow temperature range in which they have to take place as well as a possible spatial confinement of dusty clouds where the required temperature conditions for formation are fulfilled. In addition, recent observations \citep{Sakai2013,Sakai2014a,Sakai2014b,Sanna2014} have shown that the orientation of the protostellar core in the plane of the sky could be also important, particularly if it is associated to infall-disk-outflow systems and rotation.   It could be expected that the amount of chemical species resulting from the NH$_3$+O interaction will be almost equal both in high-mass and medium- to low-mass
stars  because of a higher species density in the first case and a longer interaction time in the other. All these conditions could explain the  difficulty to observe such species as hydroxylamine that have, however,  been made in the laboratory. ALMA could provide the required sensitivity and the sub-arcsecond angular resolution to observe them.  

A recent paper \citep{Cleeves2014} demonstrates, based on a physical and chemical model
of the proto-planetary disk of the solar system,
that the relative deuterated water abundance observed on Earth cannot be explained by its formation in the disk itself but should results in part
from a primeval enrichment by interstellar ices.
All species embedded in these ices may then have
survived and have been injected into the disk and
later on in the planetary system. This discovery
confirms the paramount importance of observing prebiotic species like glycine or even smaller
building blocks of such molecules like hydroxylamine, which would emphasize the possible extraterrestrial origin of life.
\section{Acknowledgments}
This work is supported by the NSF Astronomy and Astrophysics
Division (Grant No.1311958 to GV). We thank Dr. J. Brucato of the Astrophysical
Observatory of Arcetri (Italy) for providing the sample
used in these experiments and Zhirou Zhang, Jianming Shi, and Tyler Hopkins for technical help.
RTG acknowledges the support of the NASA Astrophysics Theory Program (Grant No. NNX11AC38G).

\bibliography{ref.bib}
\bibliographystyle{apj}

\end{document}